\documentclass[preprint, 3p]{elsarticle}

\usepackage{lineno,hyperref,graphicx}
\modulolinenumbers[5]

\usepackage{amssymb, amsthm, amsmath, siunitx}
\usepackage{adjustbox}

\newcommand{\UD}{{U(1)_{D}}}
\newcommand{\DF}[1]{{\xi_{(#1)}}}
\newcommand{\DFC}[1]{{\xi^{c}_{(#1)}}}

\journal{Physics of the Dark Universe}

\begin{document}

\begin{frontmatter}

%\title{Elsevier \LaTeX\ template\tnoteref{mytitlenote}}
%%% Group authors per affiliation:

\title{Anomaly-free chiral $U(1)_{D}$ and its scotogenic implication}

\author{Chi-Fong Wong}
\address{State Key Laboratory of Lunar and Planetary Sciences, Macau University of Science and Technology, Macau SAR, China}
\ead{cfwong.freeman@gmail.com}

\begin{abstract}
We assume a dark $\UD$ symmetric hidden sector and explore its possible content such that neutrino mass and Dark Matter are generated through the scotogenic mechanism.
The hidden sector is considered analogue to the Standard Model, namely the charge assignment of hidden fermions is chiral and anomaly-free.
Moreover, $\UD$ is assumed broken by an only Higgs singlet that also generates masses to all hidden fermions, therefore the charge assignment is subjected to additional restriction.
We search by computer program for charge assignments satisfying all these conditions, and identify the resulting minimal scotogenic models for Majorana neutrinos and Dirac neutrinos respectively in the sense of minimal number of messenger scalar involved.
Particle spectrum, couplings, and phenomenologies of these minimal models are briefly discussed.
DM is found multicomponent and always contains Dirac fermionic species.
\end{abstract}

\begin{keyword}
%\texttt{elsarticle.cls}\sep \LaTeX\sep Elsevier \sep template
%\MSC[2010] 00-01\sep  99-00
Exotic fermion,
Neutrino,
Dark Matter,
Anomaly cancellation
\end{keyword}

\end{frontmatter}

%\linenumbers

\section{Introduction}
\label{sec_intro}

Origins of neutrino mass and Dark Matter (DM) are two pressing puzzles in particle physics and cosmology, and are usually addressed by a hidden sector containing new particle and/or symmetry beyond the Standard Model (SM) \cite{KO:2016gxk}.
Although the content of the hidden sector is highly model-dependent, some model-independent general features may be speculated if we assume the hidden sector is analogue to the SM \cite{deGouvea:2015pea, Berryman:2016rot}, i.e.,
the hidden sector may be an anomaly-free gauge theory, in which all fermions are chiral and are massless due to gauge invariant, until spontaneous symmetry breaking (SSB) at low energy induced by vacuum expectation value (VEV) of an only Higgs multiplet.
%chiral
%gauge symmetry
%all global symmetries are accidental
%anomaly-free
%all massive fermions acquire masses from one common VEV.
The simplest realization of this picture is to assume a hidden $\UD$ symmetry.

%Roles of ACEs
Anomaly cancellation plays crucial role in this kind of models.
%It is not only required for theoretical self-consistence, but also an insightful way to understand fermion charges \cite{Geng:1988pr, Lohitsiri:2019fuu}.
It is not only required for theoretical self-consistence, but also providing an unified picture to understand individual fermion \cite{Geng:1988pr, Lohitsiri:2019fuu}.
For example, the 15 chiral fermions per generation in the SM form just the minimal solution to cancel all anomalies of the SM gauge group $SU(3)_{C} \times SU(2)_{L} \times U(1)_{Y}$.
In some SM extensions with new $U(1)$ symmetry, anomaly cancellation is employed to determine the extra particle content
%\cite{%
%%Three RHNs
%Appelquist:2002mw,
%Chen:2006hn,
%Montero:2007cd,
%%Leptoquark or leptobaryon
%Duerr:2013dza,
%Perez:2014qfa,
%%
%Hooper:2014fda,
%Kanemura:2014rpa,
%Wang:2015saa,
%Ismail:2016tod,
%Cui:2017juz,
%Ellis:2017tkh,
%Ellis:2017nrp,
%Bauer:2018egk,
%Caron:2018yzp,
%ElHedri:2018cdm,
%Allanach:2018vjg,
%Rathsman:2019wyk,
%Allanach:2020zna,
%Nakayama:2011dj,
%Nakayama:2018yvj,
%FileviezPerez:2019jju,
%FileviezPerez:2019cyn,
%Costa:2020krs}.
\cite{Appelquist:2002mw,Chen:2006hn,Montero:2007cd,Duerr:2013dza,Perez:2014qfa,Hooper:2014fda,Kanemura:2014rpa,Wang:2015saa,Ismail:2016tod,Cui:2017juz,Ellis:2017tkh,Ellis:2017nrp,Bauer:2018egk,Caron:2018yzp,ElHedri:2018cdm,Allanach:2018vjg,Rathsman:2019wyk,Allanach:2020zna,Nakayama:2011dj,Nakayama:2018yvj,FileviezPerez:2019jju,Costa:2020krs}.
This approach is greatly predictive about the number, charges, and couplings of new fermions.
Since neutral fermion and vector-like pair acquire mass without symmetry breaking, only nontrivial chiral charge assignments
%\cite{%
%Babu:2003is,
%Davoudiasl:2005ks,
%Sayre:2005yh,
%Batra:2005rh,
%Batell:2010bp,
%Heeck:2012bz,
%deGouvea:2015pea,
%Berryman:2016rot,
%Costa:2019zzy,
%Costa:2020dph}.
\cite{Babu:2003is,Davoudiasl:2005ks,Sayre:2005yh,Batra:2005rh,Batell:2010bp,Heeck:2012bz,deGouvea:2015pea,Berryman:2016rot,Costa:2019zzy,Costa:2020dph}
are relevant to SM-analog models.

%Original scotogenic models
Tininess of neutrino mass and longevity of DM are two prominent attributes that should be naturally addressed by the hidden sector.
In the so-called scotogenic mechanism, first proposed in the Ma model \cite{Ma:2006km}, both attributes are related and explained simultaneously by introducing two or three singlet Majorana fermions, a messenger scalar doublet, and an unbroken global symmetry $Z_{2}$ under which all new particles are odd.
Majorana neutrino mass is thus forbidden at tree level and is generated at one-loop level via couplings with the new fields, resulting in naturally small neutrino masses.
Meanwhile, the lightest new state is protected by the $Z_{2}$ from decaying thus becomes a stable DM candidate.
As nature of neutrinos has not been determined by current experiments yet, Dirac neutrino is also possible and corresponding scotogenic models \cite{Gu:2007ug, Farzan:2012sa} can be constructed with similar setup, except that now one needs Dirac heavy fermions in the loop instead of Majorana, an extra messenger scalar singlet besides doublet, and at least two right-handed neutrinos (RHNs).
In Majorana or Dirac case, the number of new heavy fermions that affects the resulting number of massive light neutrinos is introduced for satisfying experiments merely instead of following an underlying principle, and it does not alter the number of DM species.

% This work
In this work, we study the possible content and implications of the hidden sector with new symmetry $\UD$, if it is a SM-analogue where scotogenic mechanism is realized after SSB and thereby both neutrino mass and DM are generated.
We have considered both the cases where neutrinos are Majorana and Dirac, respectively.
The $\UD$ that prevents hidden fermion mass before symmetry breaking gives rise the question of anomaly cancellation that leads to nontrivial chiral charge assignment of these fermions, as well as constraint on coupling constant $g_{D}$.
%
%The breaking of $\UD$ is induced by only one Higgs singlet $S$, analog to the SM.
%We assume that all hidden fermions (except RHNs in Dirac case) acquire mass after SSB to avoid constraint on massless new state.
%This assumption imposes another nontrivial requirement on the charge assignment, since charge values taken to cancel anomalies could prevent formation of some necessary Yukawa couplings in some charge assignments if there is only one Higgs singlet, rendering massless new state(s).
%Therefore, the requirements on minimality of Higgs sector and massiveness of hidden particles provide additional selection rule on charge assignment besides anomaly cancellation.
Like in the SM, the breaking of $\UD$ is assumed induced by only one Higgs singlet $S$ by which all hidden fermions (except RHNs in Dirac case) acquire mass after SSB.
This setup imposes another nontrivial restriction on the charge assignment besides anomaly cancellation, since a set of charge values taken to cancel anomalies may prevent formation of all necessary Yukawa couplings with just one Higgs singlet, rendering massless new state(s).
%Therefore, the requirements on minimality of Higgs sector and massiveness of hidden particles provide additional selection rule on charge assignment besides anomaly cancellation.
%
Scotogenic mechanism is then realized in hidden sector which generates at least two heavy fermionic states at low energy charged identically under an unbroken residual symmetry of $\UD$.
This residual symmetry plays the role of the $Z_{2}$ in the Ma model \cite{Ma:2006km} to forbid tree-level neutrino mass and DM decay.
The lightest state carrying this residual charge is then stable and becomes a DM candidate.
If there are more than one residual symmetry at low energy, additional hidden particle contributes to stable relic species in the universe, resulting in multicomponent DM.
To obtain scotogenic Dirac neutrino mass, three massless chiral fermions should be included in the charge assignment as well to play the role of RHNs.

We search by computer program for hidden sectors satisfying constraints on anomaly cancellation, minimality of Higgs sector, massiveness of hidden fermions, and realization of scotogenic mechanism.
We have identified two models for Majorana neutrinos and one model for Dirac neutrinos as the minimal models in the extent of our search.
These models are regarded minimal since only the minimal number of messenger scalar fields (one doublet for Majorana case while one doublet and one singlet for Dirac case) are introduced.
All of them predict multicomponent DM.
Moreover, the number of new heavy fermions running in the neutrino mass-generating loop (thus the number of massive neutrinos) is now determined by all aforementioned constraints for charge assignment, instead of arbitrary preference.
Therefore, attributes of neutrino sector are connected to particle spectrum and interactions of hidden sector as well as DM, in a way deeper than the original scotogenic models.
These minimal models can also be regarded as examples of gauged scotogenic models (e.g. 
%\cite{%
%Nomura:2017vzp,
%Ma:2019yfo,
%Kang:2019sab,
%CentellesChulia:2019gic,
%Geng:2017foe,
%Ma:2016nnn,
%Ho:2016aye,
%Baek:2015fea,
%Ma:2013yga,
%Chang:2011kv,
%Kanemura:2011vm}
\cite{Nomura:2017vzp,Ma:2019yfo,Kang:2019sab,CentellesChulia:2019gic,Geng:2017foe,Ma:2016nnn,Ho:2016aye,Baek:2015fea,Ma:2013yga,Chang:2011kv,Kanemura:2011vm} for Majorana case and \cite{Wang:2017mcy, Han:2018zcn, Jana:2019mez, Leite:2020wjl} for Dirac case) that could modify Renormalization Group (RG) running in the Ma model \cite{Bouchand:2012dx, Merle:2015gea, Merle:2015ica, Lindner:2016kqk}.

In Sec.\ref{sec_anomaly}, we explain the details how we solve by computer to search for all anomaly-free chiral solutions within a given extent.
The result can be compared with literature and employed to different physical context.
Solutions are selected to build the minimal scotogenic models for Majorana neutrinos in Sec. \ref{sec_MModels} and for Dirac neutrinos in Sec. \ref{sec_DModels}, respectively, with basic structure and phenomenologies of these models discussed.
Conclusion is in Sec. \ref{sec_con}.
In \ref{sec_app}, we explain the algorithm used to determine whether a generic chiral fermion set gives rise inevitably massless state if only one Higgs presents.

\section{Anomaly-free chiral fermions}
\label{sec_anomaly}

%Setup
Consider a hidden sector beyond the SM contains gauge symmetry $\UD$ and $N$ SM-singlet chiral fermions $\DF{z_{i}}$ with nonzero $\UD$ charges $z_{i}$, where $i=1, \dots, N$.
These new fermions are assumed right-handed without loss of generality.
All SM fields are neutral under $\UD$.
Therefore, cancellation of gauge anomaly \cite{Adler:1969gk, Bell:1969ts, Bardeen:1969md} and mixed gauge-gravitational anomaly \cite{Delbourgo:1972xb, Eguchi:1976db, AlvarezGaume:1983ig} imposes the following constraints on charge assignment $\{\vec{z}\} = \{z_{1}, z_{2}, \dots, z_{N}\}$:
\begin{equation}\label{eq_ACEs}
\sum_{i}^{N} z_{i}^{3} = 0,
\quad
\sum_{i}^{N} z_{i} = 0.
\end{equation}
The global Witten anomaly \cite{Witten:1982fp} is not relevant since there is no new $SU(2)_{L}$ doublet.

% Requirements
%We consider all $z_{i}$ are integer rather than real, since $\UD$ is believed embedded in a nonabelian group to avoid Landau pole at some high energy \cite{Batra:2005rh}, rendering that charges must be rational numbers \cite{Slansky:1981yr, Banks:2010zn} hence rescalable to integers via redefining gauge coupling $g_{D}$.
We consider all $z_{i}$'s are integer rather than real, since abelian charge is believed rational number \cite{Slansky:1981yr, Banks:2010zn} due to embedding it in a nonabelian group to avoid Landau pole \cite{Batra:2005rh}, while rational charge can be rescaled to integer via redefining gauge coupling $g_{D}$.
%We also demand that $\{\vec{z}\}$ is chiral, i.e., no any two elements $z_{i}$ and $z_{j}$ from $\{\vec{z}\}$ satisfy $z_{i} + z_{j} = 0$, to avoid the arbitrariness on charge value $z_{i}$ ($= - z_{j}$), on number of such $\{z_{i}, z_{j}\}$ pair, and on Dirac mass scale of $\mathcal{L} \supset \overline{\DFC{z_{i}}} \DF{z_{j}}$ that is not protected by symmetry.
We also demand that $\{\vec{z}\}$ is chiral, i.e., no any pair of elements $\{z_{i}, z_{j}\}$ from $\{\vec{z}\}$ satisfy $z_{i} + z_{j} = 0$.
This is not only an analogue to the SM, and to avoid the undetermined Dirac mass scale in $\mathcal{L} \supset \overline{\DFC{z_{i}}} \DF{z_{j}}$ that is not protected by symmetry, but also to avoid the arbitrariness on value of $z_{i}$ ($= - z_{j}$) and number of such pairs.

% How to search
Solving integer solutions of Eq.\ref{eq_ACEs} is a kind of Diophantine problem and can be highly nontrivial.
Group theoretical and algebraic methods have been proposed to construct analytic solutions in terms of some free parameters \cite{Batra:2005rh, deGouvea:2015pea, Berryman:2016rot, Costa:2019zzy, Rathsman:2019wyk, Costa:2020dph}.
For phenomenological use and model-building, however, explicitly listing the solutions numerically could sometimes be more applicable.
For this sake, we build a computer program to look for integer chiral solutions of Eq.\ref{eq_ACEs}.
Given fermion number $N$, we iterate each of $z_{1}, z_{2}, \dots, z_{N}$ over integers in a prescribed interval $[-Z_{max}, Z_{max}]$ to search for occasions where $\{\vec{z}\}$ is chiral and anomaly-free.
For an acceptable running time, $5 \leq N \leq 12$ and $Z_{max} = 12$ are chosen.
The reason we ignore the cases with $N < 5$ is justified by following observations.
If $N=1$, the charge can only be zero;
if $N=2$, the charges are vector-like;
if $N=3$, there is never integer solution due to the Fermat last theorem applying on the cubic equation;
if $N=4$, the resulting charge assignment is just consisted of two vector-like pairs \cite{Davoudiasl:2005ks, Nakayama:2011dj} \footnote{
If $N=5$, it can be proved that all charges are different \cite{Davoudiasl:2005ks}.}.

\begin{figure}
\centering
\includegraphics[width=8.6cm]{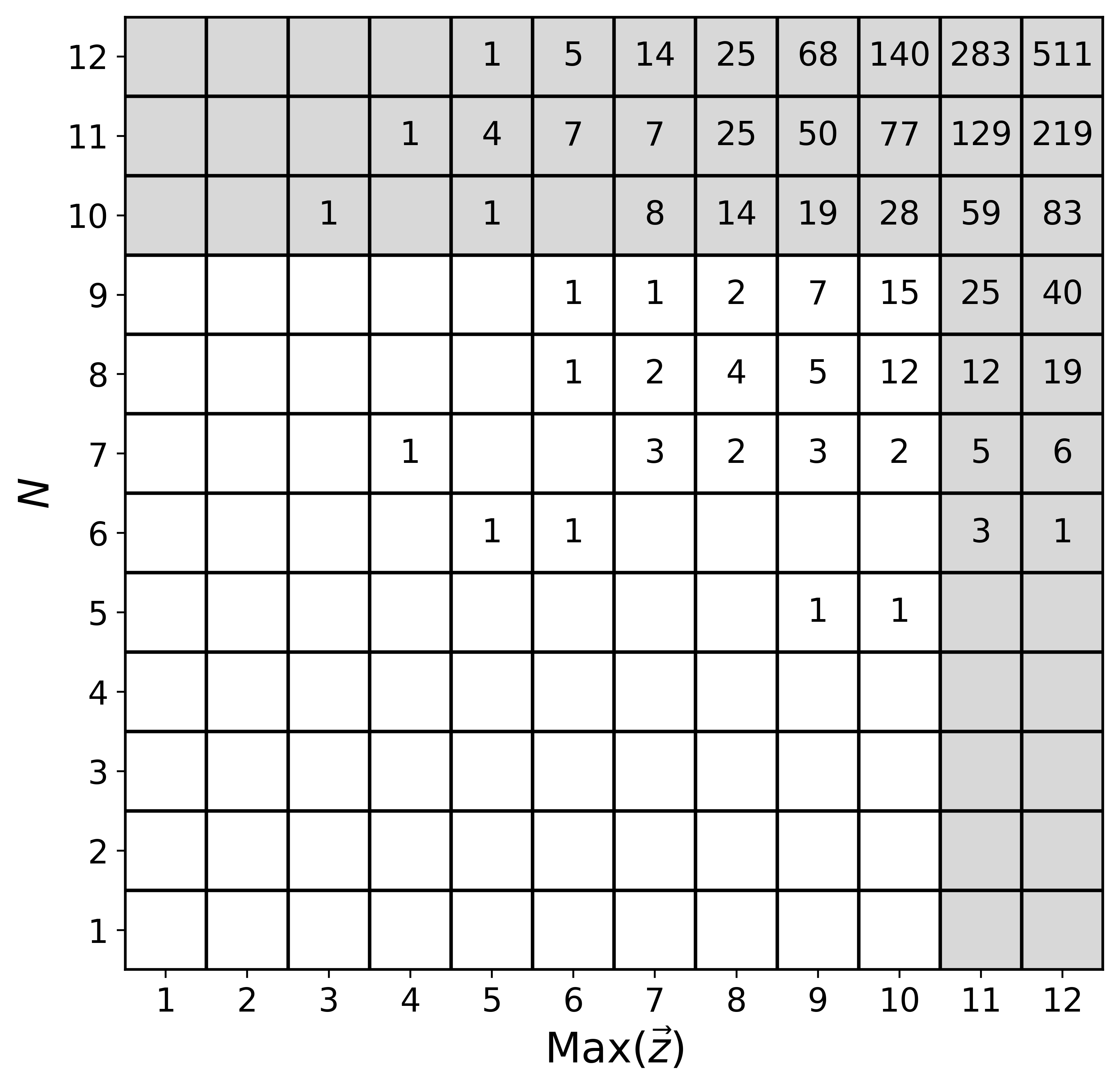}
\caption{\label{fig_birdview}
Number distribution of anomaly-free chiral charge assignments, in respect to fermion number $N$ and maximal absolute charge $\text{Max}(\vec{z})$.
Solutions in un-shadowed region are listed in TABLE \ref{table_solutions}.}
\end{figure}

% Our result
We have found 1955 primitive (i.e. coprime and non-composite \cite{Costa:2019zzy}) solutions.
%These solutions are counted according to their fermion number $N$ and maximal absolute charge $\text{Max}(\vec{z})$, giving rise the birdview shown in FIG. \ref{fig_birdview}. 
The numbers of solutions corresponding to different fermion number $N$ and maximal absolute charge $\text{Max}(\vec{z})$ are counted and shown in FIG. \ref{fig_birdview}. 
Due to limiting space, also being sufficient for our discussion hereafter, we list only solutions corresponding to $N \leq 9$ and $\text{Max}(\vec{z}) \leq 10$ explicitly in TABLE \ref{table_solutions}.
Elements in each solution are in ascending order according to their absolute values, with the first element made positive.
Some of the solutions have been explored in existing models \cite{Babu:2003is, Heeck:2012bz, deGouvea:2015pea} and similar tables \cite{Sayre:2005yh, Batra:2005rh, Nakayama:2011dj}.
Our result validates and complement to literature.
Similar computer search on anomaly-free solutions has been done in \cite{Allanach:2018vjg} with different physical context where SM fields are charged under the new symmetry and number of new particles is at most three.

\begin{table*}
%\begin{ruledtabular}
\begin{adjustbox}{width=\textwidth}
\begin{tabular}{|ccccccccc|c|c||ccccccccc|c|c||ccccccccc|c|c|}
\hline
\multicolumn{9}{|c|}{$\{\vec{z}\}$} & $z_{S}^{1}$ & $z_{S}^{2}$ & \multicolumn{9}{c|}{$\{\vec{z}\}$} & $z_{S}^{1}$ & $z_{S}^{2}$ & \multicolumn{9}{c|}{$\{\vec{z}\}$} & $z_{S}^{1}$ & $z_{S}^{2}$
\\\hline
2 & 4 & -7 & -9 & 10 &   &   &   &   &  &  & 1 & 2 & -3 & -3 & -3 & 7 & 9 & -10 &   &   &   & 1 & 1 & 2 & 2 & 3 & -5 & -6 & -6 & 8 &  & 
\\\hline 
1 & 5 & -7 & -8 & 9 &   &   &   &   &   &   & 1 & 2 & 3 & 3 & -4 & -6 & -8 & 9 &   & 5 &  & 1 & -2 & -2 & -2 & 5 & -7 & 8 & 9 & -10 &   &  
\\\hline 
1 & 1 & 1 & -4 & -4 & 5 &   &   &   &   &   & 1 & -2 & -2 & 4 & 5 & -7 & -7 & 8 &   & 9 &  & 1 & -2 & -3 & -3 & -3 & 4 & 8 & 8 & -10 &  & 
\\\hline 
1 & -2 & -3 & 5 & 5 & -6 &   &   &   &  &  & 1 & 2 & 2 & 4 & -5 & -5 & -7 & 8 &   & 3 &  & 1 & 2 & 2 & 3 & 3 & -6 & -7 & -8 & 10 &  & 
\\\hline 
1 & 1 & -3 & -6 & 8 & 9 & -10 &   &   &  &  & 1 & 2 & 2 & 2 & -3 & -5 & -6 & 7 &   & 4 &  & 1 & 2 & -3 & 4 & -5 & -6 & 8 & 8 & -9 &   &  
\\\hline 
1 & 1 & -4 & -4 & 7 & 8 & -9 &   &   &  &  & 2 & 2 & 3 & 3 & -5 & -6 & -9 & 10 &   &   &   & 1 & -2 & 3 & 4 & 6 & -7 & -7 & -7 & 9 &  & 7
\\\hline 
1 & 1 & -3 & -4 & 6 & 6 & -7 &   &   &  &  & 2 & 2 & 2 & 2 & -5 & -5 & -5 & 7 &   &   &   & 1 & 2 & 2 & 2 & 2 & -5 & -5 & -8 & 9 &  & 
\\\hline 
1 & 2 & 2 & -3 & -3 & -3 & 4 &   &   &  &  & 1 & -3 & -5 & -6 & 7 & 8 & 8 & -10 &   & 2 &  & 1 & 2 & -4 & -5 & -5 & 6 & 6 & 6 & -7 &   &  
\\\hline 
1 & 2 & -5 & -5 & 8 & 9 & -10 &   &   &  &  & 1 & -3 & -4 & -5 & 6 & 6 & 8 & -9 &   & 3 &  & 2 & 2 & -3 & 5 & -6 & 7 & -8 & -9 & 10 &  & 
\\\hline 
1 & -2 & 3 & 3 & -6 & -7 & 8 &   &   &  &  & 1 & 3 & 3 & 3 & -5 & -7 & -7 & 9 &   & 4 &  & 2 & 2 & 4 & -5 & -5 & -5 & 8 & 8 & -9 &   &  
\\\hline 
2 & 2 & -4 & 7 & -8 & -8 & 9 &   &   &   &   & 2 & -3 & -3 & 5 & 6 & -8 & -8 & 9 &   & 11 &  & 2 & 2 & 2 & -3 & -3 & 4 & -5 & -5 & 6 &   &  
\\\hline 
1 & 3 & -4 & 5 & -6 & -6 & 7 &   &   &  &  & 2 & -3 & -4 & 5 & -6 & 7 & 7 & -8 &   & 1 &  & 1 & 3 & -4 & -4 & 5 & 6 & -8 & -9 & 10 &   &  
\\\hline 
2 & 3 & -5 & 6 & -7 & -8 & 9 &   &   &  &  & 3 & -4 & -5 & 6 & -8 & 9 & 9 & -10 &   & 1 &  & 1 & -3 & 4 & 5 & 5 & -6 & -7 & -9 & 10 &   &  
\\\hline 
2 & 3 & 3 & -4 & -5 & -6 & 7 &   &   &   &   & 3 & -4 & -4 & 6 & 7 & -9 & -9 & 10 &   & 13 &  & 2 & 3 & -4 & 6 & -9 & -9 & -9 & 10 & 10 &   &  
\\\hline 
3 & 3 & 3 & -5 & -5 & -7 & 8 &   &   &   &   & 3 & -4 & -6 & -6 & 7 & 7 & 8 & -9 &   & 1 &  & 2 & 3 & -6 & -6 & 7 & 8 & -9 & -9 & 10 &   &  
\\\hline 
1 & 1 & 2 & 3 & -4 & -4 & -5 & 6 &   & 2 &  & 2 & -5 & -5 & -5 & 7 & 8 & 8 & -10 &   & 3 &  & 2 & 3 & 4 & -5 & -6 & -6 & 9 & 9 & -10 &   &  
\\\hline 
1 & 1 & 2 & -4 & 8 & -9 & -9 & 10 &   &   &   & 4 & -6 & -6 & -6 & 7 & 8 & 9 & -10 &   &   &   & 2 & -3 & -3 & -3 & -5 & 7 & 7 & 8 & -10 &   &  
\\\hline 
1 & 1 & 2 & -5 & -7 & 9 & 9 & -10 &   &   &   & 1 & 1 & -4 & -5 & 9 & 9 & 9 & -10 & -10 &  & 9 & 2 & -3 & 4 & 4 & 4 & -6 & -7 & -7 & 9 &   &  
\\\hline 
1 & 1 & 2 & -3 & 4 & -6 & -7 & 8 &   & 5 &  & 1 & 1 & 1 & 2 & -4 & 5 & -7 & -9 & 10 &   &   & 3 & 3 & -4 & 5 & 5 & -6 & -8 & -8 & 10 &   &  
\\\hline 
1 & -2 & -4 & 5 & -8 & 9 & 9 & -10 &   & 1 &  & 1 & 1 & 2 & 2 & 4 & -5 & -7 & -7 & 9 &   &   & 1 & -4 & 5 & 5 & -9 & -9 & -9 & 10 & 10 &   &  
\\\hline 
1 & 2 & 4 & 5 & -7 & -7 & -8 & 10 &   & 3 &  & 1 & 1 & 2 & 2 & 2 & -3 & -6 & -8 & 9 &   &   & 1 & 4 & 5 & -6 & -6 & -6 & 9 & 9 & -10 &   &  
\\\hline 
1 & 2 & 3 & 5 & -6 & -6 & -9 & 10 &   & 4 &  & 1 & 1 & 1 & 2 & 5 & -6 & -6 & -6 & 8 &   &   &   &   &   &   &   &   &   &   &   &   &  
\\\hline
\end{tabular}
\end{adjustbox}
\caption{\label{table_solutions}
Charge assignments of hidden sector.
Anomaly-free charge assignments for hidden fermions are shown in columns labeled by ``$\{\vec{z}\}$''.
For those in which one Higgs singlet can generate masses for all new fermions, the corresponding Higgs charges are shown in columns labeled by ``$z_{S}^{1}$''.
For those containing three right-handed neutrino candidates and the rest of new fermions are massive with one Higgs singlet, the corresponding Higgs charge are shown in columns labeled by ``$z_{S}^{2}$''.}
%\end{ruledtabular}
\end{table*}

%\subsection{Renormalization of $g_{D}$}

\begin{figure}
\centering
\includegraphics[width=8.6cm]{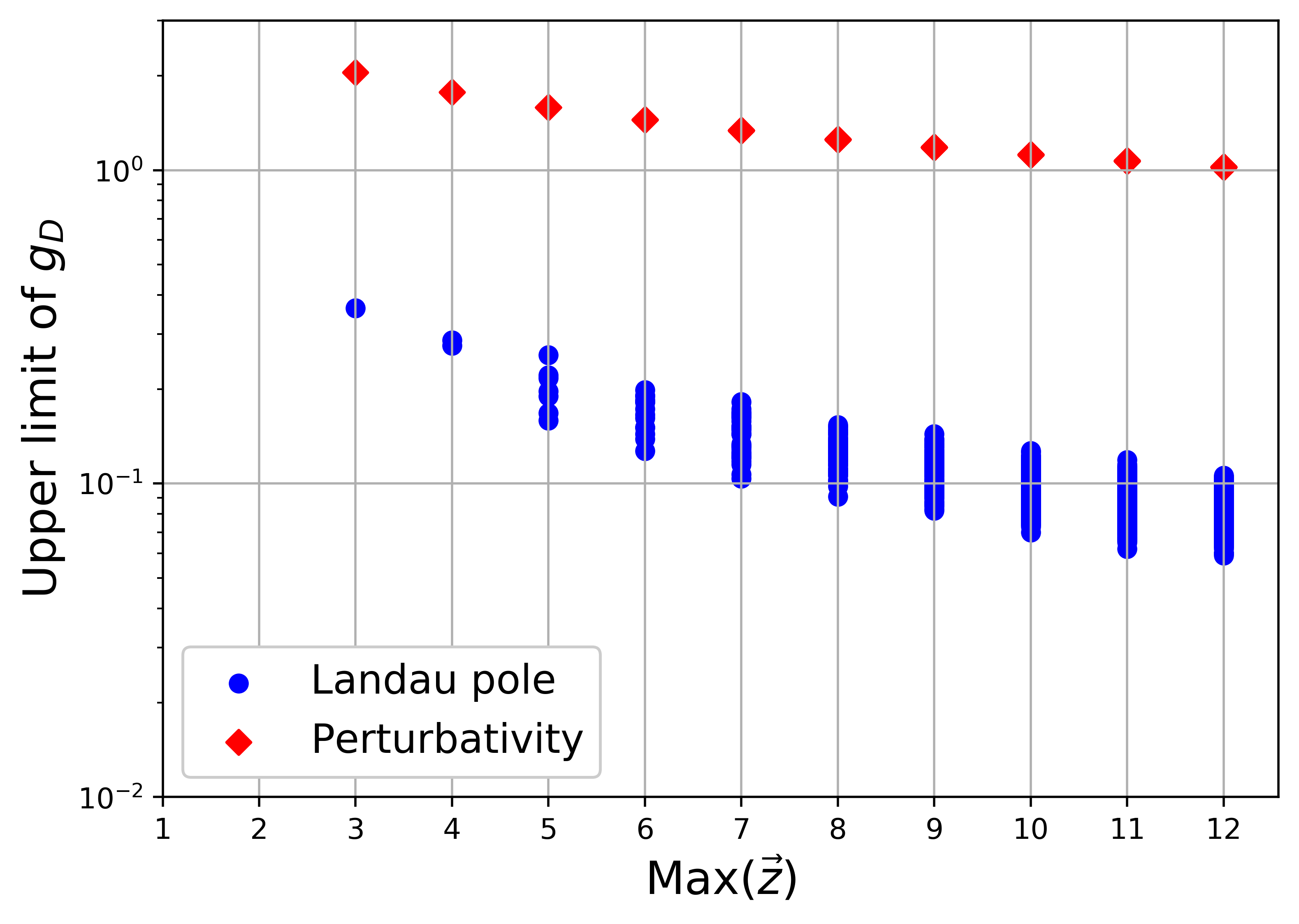}
\caption{\label{fig_gd_upper_limit}
Scatter plot of upper limit of gauge coupling $g_{D}$ at 100 GeV corresponded to each anomaly-free charge assignments imposed by Landau pole (blue points, $\Lambda_{L} = 10^{15} \si{GeV}$) and perturbativity (red diamond), respectively.}
\end{figure}

Anomaly-free conditions (Eq. \ref{eq_ACEs}) not only restrict the number and charges of new fermions, but also indirectly confine the value of gauge coupling $g_{D}$.
The presence of chiral fermions contributes to RG running of $g_{D}$.
At one-loop level, value of $g_{D}$ at energy $\mu$ is governed by $(4 \pi)^{2} d g_{D} / d \ln \mu = b g_{D}^{3}$, where $b$ is the beta function coefficient generally given by \cite{Heeck:2012bz} (see also \cite{Roy:2019jqs})
\begin{equation}
b = \frac{2}{3} \sum_{f} z_{f}^{2} + \frac{1}{3} \sum_{s} z_{s}^{2}
\end{equation}
where $z_{f}$ and $z_{s}$ the $\UD$ charges of Weyl fermionic and complex scalar DOFs respectively.
Since always $b>0$, $g_{D}$ reaches Landau pole at some energy $\Lambda_{L}$.
This provides an upper limit of $g_{D}$ at lower energy $\Lambda$:
\begin{equation}
g_{D}(\Lambda) \leq \sqrt{\frac{8 \pi^{2}}{b \ln (\Lambda_{L} / \Lambda)}}.
\end{equation}
Assuming Landau pole energy at $10^{15} \si{GeV}$, upper limits of $g_{D}$ at electroweak scale $\Lambda \sim 100 \si{GeV}$ are evaluated for each solution and are shown in FIG. \ref{fig_gd_upper_limit}.
As a comparison, perturbativity limits $g_{D} < \sqrt{4 \pi} / \text{Max}(\vec{z})$ are also given.
We can see that Landau pole limit is always stronger than perturbativity limit.
Perturbativity limit is controlled by merely the maximum charge value, while for Landau pole limit all charged DOFs contribute and that relies on determination of the whole anomaly-free chiral charge assignment.
%Larger set of new fermions prefers smaller gauge coupling.

\section{Majorana scenario}
\label{sec_MModels}

Majorana neutrino masses can be generated by scotogenic mechanism which demands two or three heavy Majorana fermions beyond the SM.
These new particles can be provided by any charge assignment in TABLE \ref{table_solutions} if one introduces enough number of Higgs singlets that develop VEVs.
However, it raises model's complexity and arbitrariness, and is also not an SM-analogue.
Instead, we look for the ``minimal'' setup in which number of scalar fields is minimized.
Namely, besides the SM Higgs doublet $\Phi$, only one Higgs singlet $S$ is introduced to break $\UD$, and only one messenger scalar doublet $\eta$ is introduced to link the SM sector with the hidden sector as in the Ma model.

%Since we assume all new fermions massive to avoid experimental constraints directly, the charge value $z_{S}$ of the sole Higgs singlet $S$ and even the existence of $S$ becomes highly nontrival.
Since we assume all new fermions massive to directly avoid further constraint from $N_\text{eff}$ measurement \cite{Aghanim:2018eyx}, the charge value $z_{S}$ of the sole Higgs singlet $S$ and even the existence of $S$ becomes highly nontrival.
For example, in some anomaly-free charge assignments, one can never obtain all necessary Yukawa couplings by just one Higgs, rendering massless fermion(s) inevitably.
We develop an algorithm to identify this kind of charge assignment and apply it to each shown in TABLE \ref{table_solutions}, to pick up those containing no inevitable massless fermion and obtain the values of $z_{S}$.
As we can see, only a small portion of charge assignments satisfy this requirement.
The algorithm is explained in \ref{sec_app}.

%Furthermore, the Majorana nature of the new heavy fermions required by scotogenic mechanism demands the value of $z_{S}$ an even number, and the neutrino oscillation data demands at least two such Majorana new fermions with charge $z_{S}/2$ (up to a sign).
Furthermore, the Majorana nature of the new heavy fermions required by scotogenic mechanism demands the value of $z_{S}$ an even number and correspondingly the Majorana fermion charge $z_{S}/2$ (up to a sign).
The neutrino oscillation data that indicate at least two nonzero neutrino masses would demand at least two such Majorana new fermions presented \cite{Cheung:2004xm}.
These observations lead to only two charge assignments from TABLE \ref{table_solutions} suitable for building scotogenic models of Majorana neutrino:
\begin{subequations}\label{eq_MModeCharges}
\begin{align}
\text{Model A:} \ & \{1, 1, 2, 3, -4, -4, -5, 6 \} \ \text{with} \ z_{S} = 2,
\label{eq_MModel1_charge}
\\
\text{Model B:} \ & \{1, 2, 2, 2, -3, -5, -6, 7 \} \ \text{with} \ z_{S} = 4.
\label{eq_MModel2_charge}
\end{align}
\end{subequations}

In what follows, we apply these charge assignments to build scotogenic models.
After working out the particle spectrum and interactions, phenomenologies will be briefly discussed, with emphasis on neutrino mass and DM physics where DM abundance is assumed thermal relics from the early universe.

\subsection{Model A}
\label{sec_MModel1}

%Particle content
The SM is extended by a set of chiral fermions $\DF{1}_{1,2}, \ \DF{2}, \ \DF{3}, \ \DF{-4}_{1,2}, \ \DF{-5}, \ \DF{6}$, with an extended scalar sector
%\begin{equation}\label{eq_MModel1_scalars}
%\begin{aligned}
%& \Phi \sim (\textbf{2}, \frac{1}{2}, 0), \quad \eta \sim (\textbf{2}, \frac{1}{2}, 1), \quad S \sim (\textbf{1}, 0, 2)
%\end{aligned}
%\end{equation}
$\Phi \sim (\textbf{2}, 1/2, 0)$,
$\eta \sim (\textbf{2}, 1/2, 1)$,
$S \sim (\textbf{1}, 0, 2)$
with their charges under $SU(2)_{L}$, $U(1)_{Y}$, and $\UD$ indicated.
% Gauge coupling saturating Landau pole limit
Beta function coefficient $b = 74$, giving rise upper limit of $g_{D}$ at 100 GeV around 0.189 (0.165) for $\Lambda_{L} = 10^{15} (10^{19}) \si{GeV}$.

%Fermion eigenstates
This particle content gives rise Yukawa couplings
\begin{equation}\label{eq_MModel1_yukawa}
\begin{aligned}
- \mathcal{L}_{Y}
= \ & Y_{aj}^{\ast} \overline{L_{a}} \ \tilde{\eta} \ \DF{1}_{j}
+ f_{ij} \ \overline{\DFC{1}_{i}} \DF{1}_{j} S^{\ast}
\\
& + h_{i} \ \overline{\DFC{2}} \DF{-4}_{i} S
+ h^{\prime}_{i} \ \overline{\DFC{6}} \DF{-4}_{i} S^{\ast}
\\
& + k \ \overline{\DFC{3}} \DF{-5} S
\\
& + h.c.
\end{aligned}
\end{equation}
where $i, j = 1, 2$, $a = 1, 2, 3$, and $\tilde{\eta} = i \sigma_{2} \eta^{\ast}$ with $\sigma_{2}$ the second Pauli matrix.
At low energy $\langle S \rangle \neq 0$, the model gives rise two Majorana fermions $N_{1,2}$, and three Dirac fermions $\Psi_{1,2}$ and $\Sigma$.
The charge 1 chiral states form Majorana fermions $N_{i} = \DFC{1}_{i} + \DF{1}_{i}$ with $i = 1, 2$, and they are responsible for generating neutrino masses.
The charge 3 and --5 chiral states merge to be Dirac fermion $\Sigma = \DFC{3} + \DF{-5}$.
The charge 2, --4, and 6 chiral states form the remaining two Dirac fermions through mass matrix
\begin{equation}\label{eq_MModel1_psiMassMatrix}
- \mathcal{L}_{Y}
\supset
\langle S \rangle
\begin{pmatrix}
\DFC{2} & \DFC{6}
\end{pmatrix}
\begin{pmatrix}
h_{1} & h_{2} \\
h_{1}^{\prime} & h_{2}^{\prime}
\end{pmatrix}
\begin{pmatrix}
\DF{-4}_{1} \\
\DF{-4}_{2}
\end{pmatrix}
+ h.c.
\end{equation}
The mass matrix can be diagonalized by biunitary transformation consisted of $U_{L}$ and $U_{R}$, giving rise mass eigenstates $\Psi_{1,2} = U_{L}^{\dagger} (\DFC{2}, \DFC{6})^{T} + U_{R}^{\dagger} (\DF{-4}_{1}, \DF{-4}_{2})^{T}$.
%Fermion interaction
These mass eigenstates interact with new gauge bosons with $\mathcal{L} \supset g_{D} X_{\mu} j^{\mu}$, where $X_{\mu}$ is the $\UD$ gauge field and
\begin{equation}\label{eq_MModel1_gaugeCurrent}
\begin{aligned}
j^{\mu}
= \ & \frac{1}{2} \overline{N_{i}} \gamma^{\mu} \gamma^{5} N_{i}
- \overline{\Sigma} \gamma^{\mu} [3 P_{L} + 5 P_{R}] \Sigma
\\
& - \begin{pmatrix}
\overline{\Psi_{1}} & \overline{\Psi_{2}}
\end{pmatrix}
\gamma^{\mu}
\left[
U_{L}^{\dagger}
\begin{pmatrix}
2 & 0 \\
0 & 6
\end{pmatrix}
U_{L}
P_{L}
+
4 P_{R}
\right]
\begin{pmatrix}
\Psi_{1} \\
\Psi_{2}
\end{pmatrix},
\end{aligned}
\end{equation}
with $P_{L}$ and $P_{R}$ are projection operators.
Charge difference of $\DF{2}$ and $\DF{6}$ results in nontrivial off-diagonal coupling between $\Psi_{1}$ and $\Psi_{2}$ therefore a hidden Flavor Changing Neutral Current (see also \cite{deGouvea:2015pea}).
As for the right-handed coupling, the identical charge of $\DF{-4}_{1}$ and $\DF{-4}_{2}$ makes $U_{R}$ vanishes and unphysical.

%Scalar sector
The most general renormalizable scalar potential contains only Hermitian operators and preserves any $U(1)$ number.
Lepton number violation is achieved by including a dimension-5 operator \cite{Kubo:2006rm}:
\begin{equation}\label{eq_MModel1_scalarPotential}
\begin{aligned}
V
= \ & - \mu_{1}^{2} |\Phi|^{2}
+ \mu_{2}^{2} |\eta|^{2}
- \mu_{3}^{2} |S|^{2}
+ \lambda_{1} |\Phi|^{4}
+ \lambda_{2} |\eta|^{4}
+ \lambda_{3} |S|^{4}
\\
& + \lambda_{12} |\Phi|^{2} |\eta|^{2}
+ \lambda_{12}^{\prime} |\Phi^{\dagger} \eta|^{2}
+ \lambda_{13} |\Phi|^{2} |S|^{2}
+ \lambda_{23} |\eta|^{2} |S|^{2}
\\
& + \frac{c}{\Lambda} (\Phi^{\dagger} \eta)^{2} S^{\ast} + h.c.
\end{aligned}
\end{equation}
with $\Lambda$ the cutoff scale \footnote{
Renormalizability can be recovered at tree level by introducing either scalar doublet $(\textbf{2}, 1/2, -1)$ or singlet $(\textbf{1}, 0, 1)$ \cite{Chang:2011kv}, that opens the dim-5 operator without generating new nonzero VEV.
We assume such additional scalar is much heavier (i.e., $\langle S \rangle \ll \Lambda$), and the operator in Eq.\ref{eq_MModel1_scalarPotential} is the only source of lepton number violation in our discussions.
%For an example of loop level, one can see \cite{}.
}.
In unitary gauge, $\Phi = (0, (v + \phi) / \sqrt{2})^{T}$, $\eta = (\eta^{+}, \eta^{0})^{T}$, and $S = (u + \phi_{S}) / \sqrt{2}$, where $v \simeq 246 \si{GeV}$ and $u \sim \mathcal{O}(\si{TeV})$.
The VEVs break electroweak symmetry and $\UD$, leaving six physical bosons: scalar $H = \sqrt{2} Re(\eta^{0})$, pseudoscalar $A = \sqrt{2} Im(\eta^{0})$, charged scalars $\eta^{\pm}$, and the two Higgs bosons $h = \cos\theta_{h} \phi + \sin\theta_{h} \phi_{S}$ and $h^{\prime} = \cos\theta_{h} \phi_{S} - \sin\theta_{h} \phi$ where $\theta_{h}$ is the rotation angle for diagonalization of mass matrix of $(\phi, \phi_{S})$:
\begin{equation}\label{eq_MModel1_higgsMassMatrix}
\begin{pmatrix}
2 \lambda_{1} v^{2}		& \lambda_{13} v u		\\
\lambda_{13} v u		& 2 \lambda_{3} u^{2}
\end{pmatrix}.
\end{equation}
The light Higgs $h$ is identified as observed at LHC with mass 125 GeV \cite{Tanabashi:2018oca}.

%Dark Matter: Residual symmetries
Spontaneous breaking of $\UD$ leaves three residual symmetries at low energy.
Besides the Krauss-Wilczek $Z_{2}$ parity \cite{Krauss:1988zc} carried by $H$, $A$, $\eta^{\pm}$, and $N_{1,2}$, there are two accidental global $U(1)$ symmetries, i.e., $U(1)_{\Psi}$ carried by $\Psi_{1,2}$ and $U(1)_{\Sigma}$ carried by $\Sigma$.
Therefore, all neutral particles carrying these residual global charges fall into three distinct Dark Sectors (DSs):
\begin{subequations}
\begin{align}
& \text{DS-1:} \quad \lbrace N_{1}, N_{2}, H, A \rbrace \\
& \text{DS-2:} \quad \lbrace \Psi_{1}, \Psi_{2} \rbrace \\
& \text{DS-3:} \quad \lbrace \Sigma \rbrace
\end{align}
\end{subequations}
The lightest state in each DS is stable.
We assume that they are generated thermally in early universe by interactions with the SM background until freeze out.
Therefore, these stable states play the role of Weakly Interacting Mass Particles (WIMPs) DM and their thermal relics together explain the observed DM abundance \cite{Aghanim:2018eyx}.

\begin{figure}
\centering
\includegraphics[width=8.6cm]{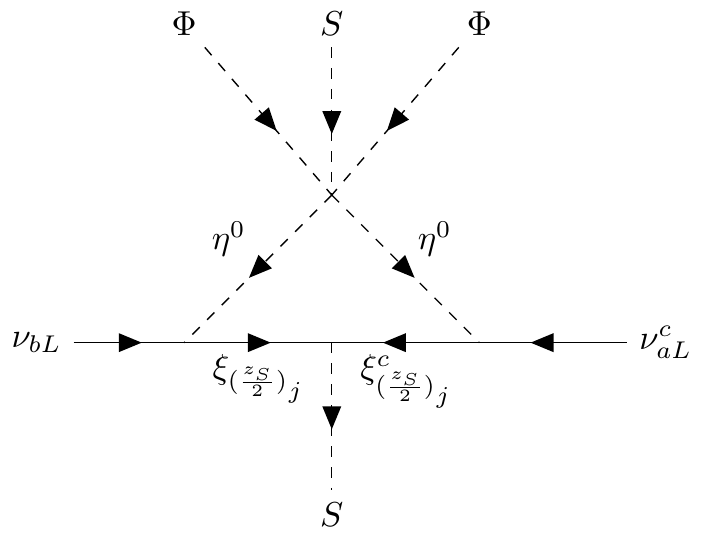}
\caption{\label{fig_MModel_nuMass}
Majorana neutrino masses generated at one loop level, where $\DF{\frac{z_{S}}{2}}$s are the Majorana DOFs carrying $\UD$ charge $z_{s}/2$.}
\end{figure}

%Neutrino mass
Particles in DS-1 are also responsible for generating scotogenic neutrino masses through one-loop diagram in FIG. \ref{fig_MModel_nuMass}.
That is \cite{Ma:2006km}
\begin{equation}\label{eq_MModel1_nuMass}
\begin{aligned}
(m_{\nu})_{ab} & = \sum_{k = 1}^{2} \frac{Y_{ak} Y_{bk} m_{N_{k}}}{16 \pi^{2}} \ \left[ \frac{m_{H^{0}}^{2}}{m_{H^{0}}^{2} - m_{N_{k}}^{2}} \ln \frac{m_{H^{0}}^{2}}{m_{N_{k}}^{2}} - \frac{m_{A^{0}}^{2}}{m_{A^{0}}^{2} - m_{N_{k}}^{2}} \ln \frac{m_{A^{0}}^{2}}{m_{N_{k}}^{2}} \right]
\end{aligned}
\end{equation}
and is proportional to difference between contributions from scalar $H$ and pseudoscalar $A$.
If we assume $H$, $A$, and $N_{1,2}$ are close at mass, and splitting between $H$ and $A$, i.e., $\delta m^{2} = m_{H}^{2} - m_{A}^{2} = 2 \sqrt{2} v^{2} c u / \Lambda$, is tiny, neutrino masses can be simplified to $(m_{\nu})_{ab} \simeq (\delta m^{2} / 32 \pi^{2}) \sum_{k = 1}^{2} (Y_{ak} Y_{bk} / m_{N_{k}})$.
Therefore $m_{\nu} \sim 0.1 \si{eV}$ can be obtained by $Y \sim 0.01$, $m_{N_{1,2}} \sim 500 \si{GeV}$, and $c u / \Lambda \sim 10^{-6}$, for example.
Since there are only two Majorana fermions involved in the loop, the lightest neutrino is massless, still consistent to experiments \cite{Ibarra:2016dlb}.
With similar parameter values, $\mu \to e \gamma$ prediction is found satisfying current limit from MEG collaboration \cite{TheMEG:2016wtm}.
Detailed discussions including other Charged Lepton Flavor Violation (CLFV) processes may be referred to \cite{Toma:2013zsa, Aoki:2016wyl, Herrero-Garcia:2016uab}.
There is also a two-loop diagram resulting in a dimension-5 operator by connecting the two $S$-legs in FIG. \ref{fig_MModel_nuMass}.
This operator, although has lower mass dimension in respect to the dimension-7 operator at one loop level in FIG. \ref{fig_MModel_nuMass}, is further suppressed by extra loop factor and propagator.
Hence we consider Eq. \ref{eq_MModel1_nuMass} the dominant form of neutrino masses.

%Dark Matter: DM1
The first DM species ($\text{DM}_{1}$) is the lightest state in DS-1 that is either $N_{1}$ or $H$ (assuming $m_{A} > m_{H}$).
Various aspects of these candidates are well accounted for in the context of Inert Doublet Model and Ma model with a large body of literature which can be found in, e.g., references in \cite{Ahriche:2017iar, Borah:2017dfn}.
For $\text{DM}_{1} = N_{1}$, DM annihilations into SM leptons in t-channel via $\overline{L_{a}} \tilde{\eta} N_{1}$ and coannihilation with $\eta$ are efficient enough if CLFV experiments tension is alleviated (e.g., \cite{Kubo:2006yx, Suematsu:2009ww, Vicente:2014wga, Ibarra:2016dlb, Ahriche:2017iar, Kitabayashi:2018bye}).
For $\text{DM}_{1} = H$, observed relic abundance can be addressed by annihilations via couplings $H H h$, $H H h h$, $H H W W$, $H H Z Z$, and coannihilation via coupling $H A Z$ and Yukawa coupling (e.g., \cite{Barbieri:2006dq, Arina:2009um, Dolle:2009fn, Honorez:2010re, LopezHonorez:2010tb, Klasen:2013jpa, Boehm:2006mi, LopezHonorez:2006gr, Borah:2017dfn, Borah:2018rca, Banerjee:2019luv, Lu:2019lok}).
%Portals
The $h$-$h^{\prime}$ mixing and $Z$-$Z^{\prime}$ mixing (induced by $\mathcal{L} \supset - \epsilon X_{\mu \nu} B^{\mu \nu} / 2$ \cite{Kumar:2006gm,Chang:2006fp,Arcadi:2017kky} where $X_{\mu}$ and $B_{\mu}$ the gauge fields of $\UD$ and $U(1)_{Y}$ respectively) do not modify much on this picture, due to the small mixings constrained by LHC and Electroweak Precision Tests \cite{Chang:2006fp} and DM direct searches \cite{Escudero:2016gzx, Arcadi:2017kky}.
General discussions on the extra gauge boson $Z^{\prime}$ can be found in \cite{Langacker:2008yv}.

%Dark Matter: DM2 and DM3
The second and the third DM species are the lightest states in DS-2 and DS-3 respectively, i.e., $\text{DM}_{2} = \Psi_{1}$ and $\text{DM}_{3} = \Sigma$.
Both DM$_{2}$ and DM$_{3}$ communicate to SM species only through mediators $Z$, $Z^{\prime}$, $h$, and $h^{\prime}$, but the parameter space has almost been ruled out by spin-independent DM-nuclei elastic scattering experiments \cite{Escudero:2016gzx, Arcadi:2017kky}.
Therefore $\text{DM}_{2}$ and $\text{DM}_{3}$ reduce their densities in early universe primarily by DM conversion $\text{DM}_{2,3} \overline{\text{DM}_{2,3}} \to \text{DM}_{1} \text{DM}_{1}$, mediated by $Z^{\prime}$ and $h^{\prime}$.
For example, thermal averaged cross section of $\Psi_{1} \overline{\Psi_{1}} \to N N$ mediated by $Z^{\prime}$ in s-channel is estimated by dimensional analysis
\begin{equation}\label{eq_MModel1_dmConversion}
\begin{aligned}
\langle \sigma v \rangle & (\Psi_{1} \overline{\Psi_{1}} \to N N)
\sim \frac{g_{D}^{4} m_{\Psi_{1}}^{2}}{m_{Z^{\prime}}^{4}}
\simeq 1 \si{pb} \times \left(\frac{m_{\Psi_{1}}}{400 \si{GeV}}\right)^{2} \left(\frac{1.4 \si{TeV}}{u}\right)^{4}
\end{aligned}
\end{equation}
where $m_{Z^{\prime}} = 2 g_{D} u$.
The resulting abundance $\Omega_{\Psi_{1}} \simeq 0.1 \si{pb} / \langle \sigma v \rangle \simeq 0.1$ is at the level consistent with observations \cite{Aghanim:2018eyx}.
Mass hierarchy $m_{\text{DM}_{2}}, m_{\text{DM}_{3}} > m_{\text{DM}_{1}}$ is necessary, otherwise the inverse conversion generates overdensity for $\text{DM}_{2}$ and $\text{DM}_{3}$.
If $m_{\Psi_{1}}$ and $m_{\Sigma}$ are heavy enough, visible signals could be observed from cosmic ray resulted from present-day annihilations $\Psi_{1} \overline{\Psi_{1}}, \Sigma \overline{\Sigma} \to N_{2} N_{2}, \eta^{\pm} \eta^{\mp}$ followed by decays $N_{2} \to N_{1} \ell \overline{\ell}, N_{1} \nu \overline{\nu}$ and $\eta^{\pm} \to H W^{\pm}, A W^{\pm}$.
DM conversion $\text{DM}_{2,3} \overline{\text{DM}_{2,3}} \to \text{DM}_{1} \text{DM}_{1}$ in regions of high DM density, e.g., Galactic Center, may generate warm/hot DM particles and modify small scale structure puzzle such as cusp-core problem \cite{Kaplinghat:2000vt}.
In this model, $\text{DM}_{2}$ and $\text{DM}_{3}$ could interact indefinitely weak with quarks and leptons if the $h$-$h^{\prime}$ and $Z$-$Z^{\prime}$ mixings are too small, leaving no signal in direct detections and even in colliders.
Therefore pair productions of messenger scalar bosons $\eta^{\pm}$, $H$ and $A$ could become the most promising way to explore these new particles in colliders \cite{Ahriche:2017iar} (see also \cite{Hessler:2016kwm}).

\subsection{Model B}
\label{sec_MModel2}

%Particle content
The SM is extended by a set of chiral fermions $\DF{1}, \ \DF{2}_{1,2,3}, \ \DF{-3}, \ \DF{-5}, \ \DF{-6}, \ \DF{7}$ with an extended scalar sector
%\begin{equation}\label{eq_MModel2_scalars}
%\begin{aligned}
%& \Phi \sim (\textbf{2}, \frac{1}{2}, 0), \quad \eta \sim (\textbf{2}, \frac{1}{2}, 2), \quad S \sim (\textbf{1}, 0, 4).
%\end{aligned}
%\end{equation}
$\Phi \sim (\textbf{2}, 1/2, 0)$,
$\eta \sim (\textbf{2}, 1/2, 2)$,
$S \sim (\textbf{1}, 0, 4)$.
% Gauge coupling saturating Landau pole limit
Beta function coefficient $b = 96$, giving rise upper limit of $g_{D}$ at 100 GeV around 0.166 (0.145) for $\Lambda_{L} = 10^{15} (10^{19}) \si{GeV}$.

%Fermion eigenstates
Yukawa sector can be determined as
\begin{equation}\label{eq_MModel2_yukawa}
\begin{aligned}
- \mathcal{L}_{Y}
= \ & Y_{aj}^{\ast} \overline{L_{a}} \ \tilde{\eta} \ \DF{2}_{j}
+ f_{ij} \ \overline{\DFC{2}_{i}} \DF{2}_{j} S^{\ast}
+ f_{i}^{\prime} \ \overline{\DFC{-6}} \DF{2}_{i} S
\\
& + k \ \overline{\DFC{1}} \DF{-5} S
+ g \ \overline{\DFC{-3}} \DF{7} S^{\ast}
\\
& + h.c.
\end{aligned}
\end{equation}
where $i, j = 1, 2$, $a = 1, 2, 3$.
At low energy $\langle S \rangle \neq 0$, the model gives rise four Majorana fermions $N_{1,2,3,4}$, and two Dirac fermions $\Psi$ and $\Sigma$.
The charge 1 and --5 chiral states and charge --3 and 7 chiral states, merge to be Dirac fermions $\Psi = \DFC{1} + \DF{-5}$ and $\Sigma = \DFC{-3} + \DF{7}$ respectively.
The charge 2 and --6 chiral states mix through mass matrix
\begin{equation}\label{eq_MModel2_nMassMatrix}
\mathbb{M} = 
\begin{pmatrix}
0 & f^{\prime}_{1} & f^{\prime}_{2} & f^{\prime}_{3} \\
f^{\prime}_{1} & f_{11} &  f_{12} &  f_{13} \\
f^{\prime}_{2} & f_{12} &  f_{22} &  f_{23} \\
f^{\prime}_{3} & f_{13} &  f_{23} &  f_{33} \\
\end{pmatrix}.
\end{equation}
The mass matrix is symmetric, and can be diagonalized by unitary transformation $U$ on basis $(\DF{-6}, \DF{2}_{1}, \DF{2}_{2}, \DF{2}_{3})$, resulting in four Majorana fermions $(N_{1}, N_{2}, N_{3}, N_{4})^{T} = U^{\dagger} (\DF{-6}, \DF{2}_{1}, \DF{2}_{2}, \DF{2}_{3})^{T} + U^{T} (\DFC{-6}, \DFC{2}_{1}, \DFC{2}_{2}, \DFC{2}_{3})^{T}$.
Interestingly, the structure of $\mathbb{M}$ implies a possible Seesaw-like hierarchy among $N_{i}$, relevant to setup like \cite{Baumholzer:2018sfb,Baumholzer:2019twf}.
For example, a slight ratio $f^{\prime} / f \simeq 0.1$ could result in two order of magnitude hierarchy between $N_{1}$ and the heavier ones, i.e., $m_{N_{1}} \sim (f^{\prime} / f)^{2} m_{N_{2,3,4}} \simeq 10^{-2} m_{N_{2,3,4}}$.

%Fermion interaction
The new gauge interaction current in $\mathcal{L} \supset g_{D} X_{\mu} j^{\mu}$ contributed by $N_{i}$, $\Psi$, and $\Sigma$ reads
\begin{equation}\label{eq_MModel_gaugeCurrent}
\begin{aligned}
& j^{\mu} = \overline{\Psi} [(3) P_{L} + (7) P_{R}] \Psi
+ \overline{\Sigma} [(-1) P_{L} + (-5) P_{R}] \Sigma
\\
& + 
\frac{1}{2}
\begin{pmatrix}
\overline{N_{1}} &
\overline{N_{2}} &
\overline{N_{3}} &
\overline{N_{4}}
\end{pmatrix}
\gamma^{\mu}
\gamma^{5}
%\left[
U^{\dagger}
\begin{pmatrix}
-6 & 0 & 0 & 0 \\
0 & 2 & 0 & 0 \\
0 & 0 & 2 & 0 \\
0 & 0 & 0 & 2 \\
\end{pmatrix}
U
%\right]
\begin{pmatrix}
N_{1} \\
N_{2} \\
N_{3} \\
N_{4}
\end{pmatrix}.
\end{aligned}
\end{equation}
The 4-by-4 unitary matrix $U$ contains 16 parameters, i.e., six rotation angles and 10 phases.
Some of these parameters are unphysical.
Since the charge matrix $\text{diag}(-6, 2, 2, 2)$ is diagonal and its 3-by-3 lower-right submatrix is proportional to identity, four phases and three rotation angles are always cancelled.
Finally, only nine parameters (three angles, three Dirac-like phases, and three Majorana-like phases) are present in Lagrangian.

%Scalar sector
The scalar sector of this model gives rise the same scalar potential (Eq.\ref{eq_MModel1_scalarPotential}), physical boson spectrum, and scalar couplings as in previous model.
A distinction is of gauge interaction since now $\eta$ and $S$ carry double charges under $\UD$ in respect to those in the model A.

%Dark Matter: Residual symmetries
There are again three residual symmetries left in low energy theory.
They are $Z_{2}$ parity carried by $H$, $A$, $\eta^{\pm}$, and $N_{1,2,3,4}$; $U(1)_{\Psi}$ carried by $\Psi$; and $U(1)_{\Sigma}$ carried by $\Sigma$.
All neutral new particles fall into three DSs:
\begin{subequations}
\begin{align}
& \text{DS-1:} \quad \lbrace N_{1}, N_{2}, N_{3}, N_{4}, H, A \rbrace \\
& \text{DS-2:} \quad \lbrace \Psi \rbrace \\
& \text{DS-3:} \quad \lbrace \Sigma \rbrace
\end{align}
\end{subequations}
%Neutrino mass
%LFV decays
Neutrino masses and CLFV $\mu \to e \gamma$ decay are induced by particles in DS-1, similar to Majorana model A (Sec.\ref{sec_MModel1}).
However, number of loop fermions is now modified to four, thus all three active neutrinos can be massive.

%Dark Matter: DM1, DM2, and DM3
The three DM components are similar to those in Majorana model A: $\text{DM}_{1} = N_{1}$ or $H$; $\text{DM}_{2} = \Psi$; and $\text{DM}_{3} = \Sigma$.
DM physics is then qualitatively similar to previous model, with different scattering strength due to different $\UD$ charges carried by the new particles, and apparent different number of particles in each DS.

\section{Dirac scenario}
\label{sec_DModels}

Dirac neutrino masses can be generated in scotogenic models \cite{Gu:2007ug, Farzan:2012sa} that introduce two or three heavy Dirac fermions $\Psi_{i}$, and three chiral fermions $\nu_{Ri}$ as RHNs.
The messenger scalar doublet $\eta$ and singlet $\sigma$ mediate one loop neutrino masses by couplings $\overline{L_{i}} \tilde{\eta} \Psi_{jR}$ and $\overline{\nu_{Ri}} \sigma \Psi_{jL}$ where $L_{i}$ the SM left-handed lepton doublets, therefore the left- and right-handed part of neutrinos are connected and Dirac neutrino masses are obtained.

We look for charge assignments from TABLE \ref{table_solutions} by which $\Psi_{i}$ and $\nu_{Ri}$ can present given a sole Higgs singlet $S$, and similar to Majorana case (Sec.\ref{sec_MModels}) that all new fermions except $\nu_{Ri}$ are massive.
Such pattern of particle spectrum happens only in two solutions in TABLE \ref{table_solutions}, they are $\{\vec{z}_{1}\} = \{ 1, 1, -4, -5, 9, 9, 9, -10, -10 \}$ and $\{\vec{z}_{2}\} = \{ 1, -2, 3, 4, 6, -7, -7, -7, 9 \}$.
However, if we consider only the minimal number of messegner scalars, i.e., only $\eta$ and $\sigma$ are introduced, $\{\vec{z}_{2}\}$ can generate only one Dirac neutrino mass that violates observations, leaving only one solution suitable for building scotogenic model of Dirac neutrino:
\begin{equation}\label{eq_DModel1_charge}
\text{The model:} \ \{ 1, 1, -4, -5, 9, 9, 9, -10, -10 \} \ \text{with} \ z_{S} = 9.
\end{equation}
In what follows, we discuss the basic structure of the model and its phenomenologies.

\subsection{The model}
\label{sec_DModel1}

%Particle content
The SM is extended by a set of chiral fermions $\DF{1}_{1,2}, \ \DF{-4}, \ \DF{-5}, \ \DF{-9}_{1,2,3}, \ \DF{-10}_{1,2}$
with an extended scalar sector
%\begin{equation}\label{eq_DModel1_scalars}
%\begin{aligned}
%& \Phi \sim (\textbf{2}, \frac{1}{2}, 0), \quad \eta \sim (\textbf{2}, \frac{1}{2}, 1), \quad S \sim (\textbf{1}, 0, 9), \quad \sigma \sim (\textbf{1}, 0, 1).
%\end{aligned}
%\end{equation}
$\Phi \sim (\textbf{2}, 1/2, 0)$,
$\eta \sim (\textbf{2}, 1/2, 1)$,
$S \sim (\textbf{1}, 0, 9)$,
$\sigma \sim (\textbf{1}, 0, 1)$
\footnote{
An alternative charge assignment can be taken, in which $\eta \sim (\textbf{2}, \frac{1}{2}, -10)$ and $\sigma \sim (\textbf{1}, 0, -10)$.
Such assignment brings in 10-times larger gauge interactions for these fields.
}.
% Gauge coupling saturating Landau pole limit
Beta function coefficient $b = 352$, giving rise upper limit of $g_{D}$ at 100 GeV around 0.087 (0.076) for $\Lambda_{L} = 10^{15} (10^{19}) \si{GeV}$.

%Fermion eigenstates
Yukawa sector reads:
\begin{equation}
\begin{aligned}
- \mathcal{L}_{Y}
= \ & Y_{aj}^{\ast} \overline{L_{a}} \ \tilde{\eta} \ \DF{1}_{j}
+ K_{aj} \overline{\DF{9}_{a}} \sigma^{\ast} \DFC{-10}_{j}
\\
& + h_{ij} \overline{\DFC{1}_{i}} \DF{-10}_{j} S
+ k \overline{\DFC{-4}} \DF{-5} S
\\
& + h.c.
\end{aligned}
\end{equation}
where $i, j = 1,2$ and $a = 1,2,3$.
The 2-by-2 matrix $h_{ij}$ can be taken diagonal by redefining $\DF{1}_{i}$ and $\DF{-10}_{i}$ without loss of generality.
At low energy $\langle S \rangle \neq 0$, three Dirac fermions and three Weyl fermions are generated.
Charge 1 and --10 chiral states and charge --4 and --5 chiral states form Dirac fermions $\Psi_{i} = \DFC{1}_{i} + \DF{-10}_{i}$ with $i=1,2$ and $\Sigma = \DFC{-4} + \DF{-5}$ respectively.
The three charge 9 chiral states do not receive mass from $\langle S \rangle$ and are identified as RHNs, i.e., $\nu_{aR} = \DF{9}_{a}$.
%
%Fermion interaction
Accordingly, gauge interactions of new fermions are described by the current
\begin{equation}\label{eq_DModel1_gaugeCurrent}
\begin{aligned}
j^{\mu} = \ & \overline{\Psi_{i}} \gamma^{\mu} [(-1) P_{L} + (-10) P_{R}] \Psi_{i}
\\
& + \overline{\Sigma} \gamma^{\mu} [(4) P_{L} + (-5) P_{R}] \Sigma
\\
& + \overline{\nu_{a}} \gamma^{\mu} (9) P_{R} \nu_{a}.
\end{aligned}
\end{equation}

%Scalar sector
The most general renormalizable scalar potential \footnote{
The scalar potential has also studied in Ref. \cite{Wang:2017mcy}, up to a term $|\Phi^{\dagger} \eta|^{2}$.
}
\begin{equation}\label{eq_DModel1_scalarPotential}
\begin{aligned}
V
= \ & - \mu_{1}^{2} |\Phi|^{2}
+ \mu_{2}^{2} |\eta|^{2}
- \mu_{3}^{2} |S|^{2}
+ \mu_{4}^{2} |\sigma|^{2}
\\
& + \lambda_{1} |\Phi|^{4}
+ \lambda_{2} |\eta|^{4}
+ \lambda_{3} |S|^{4}
+ \lambda_{4} |\sigma|^{4}
\\
& + \lambda_{12} |\Phi|^{2} |\eta|^{2}
+ \lambda_{12}^{\prime} |\Phi^{\dagger} \eta|^{2}
+ \lambda_{13} |\Phi|^{2} |S|^{2}
+ \lambda_{14} |\Phi|^{2} |\sigma|^{2}
\\
& + \lambda_{23} |\eta|^{2} |S|^{2}
+ \lambda_{24} |\eta|^{2} |\sigma|^{2}
+ \lambda_{34} |S|^{2} |\sigma|^{2}
\\
& + \kappa \Phi^{\dagger} \eta \sigma^{\ast}
+ h.c.
\end{aligned}
\end{equation}
has the last term non-Hermitian.
Its coefficient $\kappa$ is expected small, since the operator breaks a global symmetry $U(1)_{\Phi} \times U(1)_{\eta} \times U(1)_{S} \times U(1)_{\sigma}$ down to $U(1)_{\Phi + 2 \eta + \sigma} \times U(1)_{S}$ in scalar potential \cite{tHooft:1979rat}.
In unitary gauge, $\Phi = (0, (v + \phi) / \sqrt{2})^{T}$, $\eta = (\eta^{+}, \eta^{0})^{T}$, $S = (u + \phi_{S}) / \sqrt{2}$, $\sigma = \sigma$.
After breaking of electroweak symmetry and $\UD$, global symmetry in the scalar potential is further broken down to $U(1)_{\eta + \sigma}$.
The presence of this symmetry implies that the real and imaginary parts in $\eta^{0}$ and $\sigma$ respectively are degenerate, giving rise complex scalar mass eigenstates.
Finally there are eight physical bosons: charged bosons $\eta^{\pm}$, complex neutral scalar bosons $\phi_{1} = \cos\theta_{\phi} \eta^{0} + \sin\theta_{\phi} \sigma$ and $\phi_{2} = \cos\theta_{\phi} \sigma - \sin\theta_{\phi} \eta^{0}$ with mixing angle $\theta_{\phi}$ diagonalizing mass matrix of $(\eta^{0}, \sigma)$:
\begin{equation}
\begin{pmatrix}
2 \mu_{2}^{2} + (\lambda_{12} + \lambda_{12}^{\prime}) v^{2} + \lambda_{23} u^{2}
&
\sqrt{2} \kappa v
\\
\sqrt{2} \kappa v
&
2 \mu_{4}^{2} + \lambda_{14} v^{2} + \lambda_{34} u^{2}
\end{pmatrix},
\end{equation}
and the two Higgs bosons $h$ and $h^{\prime}$ defined exactly as in Majorana model A thanks to the same mass matrix given in Eq.\ref{eq_MModel1_higgsMassMatrix}.

%Dark Matter: Residual symmetry
Spontaneous breaking of $\UD$ leaves two residual symmetries at low energy.
Both of them are continuous symmetries.
They are $U(1)_{\Psi}$ carried by $\phi_{1,2}$, $\eta^{\pm}$, and $\Psi_{1,2}$; and $U(1)_{\Sigma}$ carried by $\Sigma$.
Neutral particles carrying these residual charges fall into two DSs:
\begin{subequations}
\begin{align}
& \text{DS-1:} \quad \lbrace \Psi_{1}, \Psi_{2}, \phi_{1}, \phi_{2} \rbrace \\
& \text{DS-2:} \quad \lbrace \Sigma \rbrace
\end{align}
\end{subequations}

\begin{figure}
\centering
\includegraphics[width=8.6cm]{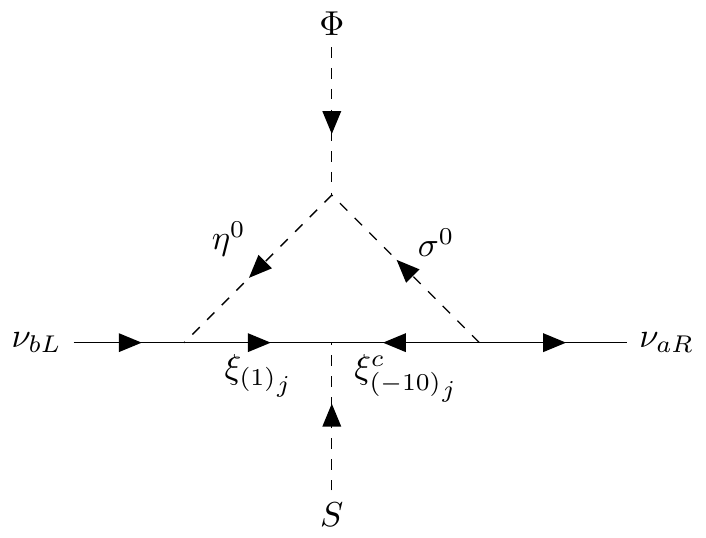}
\caption{\label{fig_DModel_nuMass}
Dirac neutrino masses generated at one loop level.}
\end{figure}

%Neutrino mass
Neutrino masses are generated by the one-loop diagram in FIG.\ref{fig_DModel_nuMass}, calculated by
\begin{equation}
\begin{aligned}
(m_{\nu})_{ab} & = \sum_{k=1}^{2} \frac{K_{ak} Y_{bk} m_{\Psi_{k}}}{16 \pi^{2}} \sin 2\theta_{\phi} \ \left[ \frac{m_{\phi^{0}_{1}}^{2}}{m_{\phi^{0}_{1}}^{2} - m_{\Psi_{k}}^{2}} \ln \frac{m_{\phi^{0}_{1}}^{2}}{m_{\Psi_{k}}^{2}} - \frac{m_{\phi^{0}_{2}}^{2}}{m_{\phi^{0}_{2}}^{2} - m_{\Psi_{k}}^{2}} \ln \frac{m_{\phi^{0}_{2}}^{2}}{m_{\Psi_{k}}^{2}} \right]
\end{aligned}
\end{equation}
Unlike Majorana models, neutrino masses are not proportional to difference between contributions from real and imaginary parts of the in-loop scalar, but between contributions from mass eigenstates of the complex scalars $\phi$'s.
Since imaginary parts of both $\eta^{0}$ and $\sigma$ contribute, there is an implicit factor of two in above expression, distinct from Ref.\cite{Farzan:2012sa} where $\sigma$ is a real scalar thus only $Re(\eta^{0})$ is involved in the loop.
If $\phi_{1,2}$ and $\Psi_{1,2}$ are close in mass and the splitting $\delta m^{2} = m_{\phi_{1}}^{2} - m_{\phi_{2}}^{2} = 2 \sqrt{2} \kappa v / \sin 2\theta_{\phi}$ is small, we have $(m_{\nu})_{ab} \simeq (\delta m^{2} / 32 \pi^{2}) \sum_{i = 1}^{2} (K_{ak} Y_{bk} \sin 2\theta_{\phi} / m_{\Psi_{k}})$ and $m_{\nu} \sim 0.1 \si{eV}$ can be obtained if $K \sim Y \sim 0.01$, $m_{\Psi_{k}} \sim 500 \si{GeV}$, and $\kappa / v \sim 10^{-6}$.
As in Majorana model A (Sec.\ref{sec_MModel1}), only two heavy fermions are involved in the loop diagram (FIG.\ref{fig_DModel_nuMass}), therefore the lightest neutrino is massless.
%LFV decays
Yukawa coupling $Y_{ai}$ induces CLFV $\mu \to e \gamma$ decay in the same way as Majorana models.

%Dark Matter: DM1
The first DM species ($\text{DM}_{1}$) is the lightest state in DS-1 which is either $\Psi_{1}$ or $\phi_{1}$.
For $\text{DM}_{1} = \Psi_{1}$, Yukawa coupling $K_{ai}$ could mediate $\Psi_{1} \overline{\Psi_{1}} \to \nu_{aR} \overline{\nu_{bR}}$ in t-channel, without constraint from CLFV experiments \cite{Farzan:2012sa}.
For $\text{DM}_{1} = \phi_{1}$, the $\eta^{0}$ component is severely suppressed due to direct detection constraints on $\eta^{0} \eta^{0 \ast} Z$ coupling \cite{Barbieri:2006dq}, thus $\phi_{1}$ is dominated by $\sigma$ and $\theta_{\phi} \simeq \pi/2$.
Therefore $\phi_{1}$ can annihilate into RHNs in t-channel mediated by $\Psi_{1,2}$, with cross section \cite{Berlin:2014tja}
\begin{equation}
\langle \sigma v_{rel} \rangle
\simeq v_{rel}^{2} \sum_{k = 1}^{2} \frac{|K_{ik}^{\ast} K_{jk}|^{2}}{96 \pi} \frac{m_{\phi_{1}}^{2} \sin^{4} \theta_{\phi}}{(m_{\phi_{1}}^{2} + m_{\Psi_{k}}^{2})^{2}}
\end{equation}
Take $k \equiv (|K_{ik}^{\ast} K_{jk}|^{2})^{1/4} \sim 0.8$, $m_{\phi_{1}} \sim 500 \si{GeV}$, $v_{rel} \sim 0.3$, $\theta_{\phi} \sim \pi / 2$, and adding up all final state flavors, we obtain $\langle \sigma v_{rel} \rangle \sim 1 \si{pb}$ as $m_{\Psi_{k}} \simeq 540 \si{GeV}$.
%Dark Matter: DM2
The second DM species $\Sigma$ from DS-2 communicates with SM by $Z$, $Z^{\prime}$, $h$, and $h^{\prime}$, but all these mediators are ineffective similar to Majorana models.
Therefore DM conversion $\Sigma \overline{\Sigma} \to \text{DM}_{1} \text{DM}_{1}$ mediated by $Z^{\prime}$ is relevant.
Indeed, even if $m_{\Sigma} < m_{\text{DM}_{1}}$, $\Sigma$ would not be overdensity in the early universe because gauge boson $Z^{\prime}$ could mediate annihilation $\Sigma \overline{\Sigma} \to \nu_{aR} \overline{\nu_{bR}}$, with cross section similar to that estimated in Eq.\ref{eq_MModel1_dmConversion}.

%Dark Radiation
RHNs are nearly massless and could play the role of dark radiation, hence the decoupling temperature of RHNs from SM plasma is constrained \cite{Cui:2017ytb}.
In this model, RHNs communicate with SM species via $\nu_{R} \sigma \leftrightarrow \nu_{L} \eta$ in s-channel and $\nu_{R} \eta^{\ast} \leftrightarrow \nu_{L} \sigma^{\ast}$ in t-channel, both mediated by $\Psi_{i}$.
The lowest possible freeze-out temperature of these scatterings must be around mass scale of $\eta$ and $\sigma$, i.e., about $\mathcal{O}(100 \si{GeV})$.
Therefore we conclude that the model is consistent with the measured $\Delta N_\text{eff}$ \cite{Aghanim:2018eyx}.

\section{Conclusion}
\label{sec_con}

%This study aims to reveal the general outcome in neutrino and DM sector if the scotogenic idea is the underlying mechanism of these puzzles and the new degrees of freedom are from a $\UD$-symmetric chiral anomaly-free hidden sector with minimalized scalar content.
This study aims to reveal the general outcome if neutrino mass and DM are originated from scotogenic mechanism while the new degrees of freedom come from a $\UD$-symmetric anomaly-free chiral hidden sector.
The massiveness of hidden fermions at low energy gives rise additional constraint due to only one Higgs singlet introduced.
%We speculate that both neutrino mass and DM in the universe are originated through scotogenic mechanism from a SM-analogue hidden sector with a new gauge symmetry $\UD$.
%The charge assignment of such hidden sector is thus constrained to be chiral and anomaly-free, also the massiveness of hidden fermions at low energy gives rise additional constraint due to only one Higgs singlet introduced.
%
We have search by computer program for charge assignments satisfying all these conditions.
The charge assignments discovered in this way allow us to investigate the RG running of the new gauge coupling constant $g_{D}$.
To avoid Landau pole, value of $g_{D}$ at electroweak scale has been restricted more stringent than considering only perturbativity.
With a minimalized messenger scalar sector, minimal models for either Majorana or Dirac neutrinos are identified.
These models show that the DM is multicomponent and always contains Dirac fermionic species.
The numbers of massive neutrinos and DM component are now related by the underlying charge assignment instead of put in by hand.
Therefore, neutrino sector and DM sector are related in a way deeper than in the original scotogenic models.

Finally, it is worth to note that more scotogenic models can be built from charge assignments in TABLE \ref{table_solutions} if one releases the restriction on the size of scalar sector.
For example, models of Majorana neutrinos with only one heavy in-loop fermion \cite{Hehn:2012kz} (see also \cite{Escribano:2020iqq}), and models with nature of the in-loop heavy fermions opposite to that of neutrinos \cite{Ma:2013yga, Calle:2019mxn}.

\section*{Acknowledgments}
This work was supported by The Science and Technology Development Fund (FDCT) of Macau SAR [grant number 0045/2018/AFJ].
Feynman diagrams in this paper are drawn by Tikz-Feynman package \cite{Ellis:2016jkw}.

\appendix  % Use 'appendix' if more than one appendix
\section{Finding out all possible Higgs sectors}
\label{sec_app}

Consider a simple case with two chiral fermions $\DF{a}$, $\DF{b}$ carrying $\UD$ charge $a$ and $b$ respectively.
%and Higgs singlet $S$ carrying $\UD$ charges $a$, $b$, and $q > 0$ respectively.
Mass term $\overline{\DFC{a}} \DF{b}$ presents in Lagrangian after SSB only if a Higgs singlet $S_{(q)}$ with charge $q = |a+b|$ exists, i.e., Kronecker delta $\delta_{a+b, \pm q} = 1$ implies the mass term exists while $=0$ implies contrary.
If there exist $K$ Higgs singlets carrying positive charges $\{\vec{q}\} = \{q_{1}, \dots, q_{K}\}$, the Kronecker delta is promoted to $\Delta^{\vec{q}}_{a+b} \equiv \delta_{a+b, \pm q_{1}} + \delta_{a+b, \pm q_{2}} + \dots + \delta_{a+b, \pm q_{K}}$, where $\Delta^{\vec{q}}_{a+b} = 1$ implies $|a+b| \in \{\vec{q}\}$ and the mass term exists while $\Delta^{\vec{q}}_{a+b} = 0$ implies contrary.

Consider a generic chiral fermion set $\DF{z_{1}}, \dots, \DF{z_{N}}$.
If all fermions are massive, the determinant of the $N\times N$ mass matrix
\begin{equation}\label{eq_grandMassMatrix}
\mathbf{M} = 
\begin{pmatrix}
m_{11} \Delta^{\vec{q}}_{z_{1}+z_{1}} & \dots & m_{1N} \Delta^{\vec{q}}_{z_{1}+z_{N}} \\
\vdots & \ddots & \vdots \\
m_{N1} \Delta^{\vec{q}}_{z_{N}+z_{1}} & \dots & m_{NN} \Delta^{\vec{q}}_{z_{N}+z_{N}}
\end{pmatrix},
\end{equation}
\begin{equation}\label{eq_grandMassMatrixDet}
\begin{aligned}
\text{Det}(\mathbf{M})
= \ & \sum_{P} m_{1P(1)} m_{2P(2)} \dots m_{NP(N)}
\\
& \times \Delta^{\vec{q}}_{z_{1}+z_{P(1)}}
\Delta^{\vec{q}}_{z_{2}+z_{P(2)}}
\dots
\Delta^{\vec{q}}_{z_{N}+z_{P(N)}},
\end{aligned}
\end{equation}
must be nonzero, where the mass matrix is written in basis $(\DF{z_{1}}, \dots, \DF{z_{N}})$, $m_{ij}$'s the general mass matrix elements being the product of Yukawa coupling and VEV, and $P(i)$ a dummy permutation on index $i$.
Nonzero $\text{Det}(\mathbf{M})$ implies at least one term in the summation $\sum_{P}$ in Eq.\ref{eq_grandMassMatrixDet} is nonzero, where we have $\Delta^{\vec{q}}_{z_{1}+z_{P(1)}} = \Delta^{\vec{q}}_{z_{2}+z_{P(2)}} = \dots = \Delta^{\vec{q}}_{z_{N}+z_{P(N)}} = 1$ or $|z_{1}+z_{P(1)}|, |z_{2}+z_{P(2)}|, \dots, |z_{N}+z_{P(N)}| \in \{q_{1}, q_{2}, \dots, q_{K} \}$.
In other words, Higgs sector that generates masses to all chiral fermions can be constructed by following charge assignment
\begin{equation}
\{ |z_{1}+z_{P(1)}|, |z_{2}+z_{P(2)}|, \dots, |z_{N}+z_{P(N)}| \}
\end{equation}
once the $P$ is chosen.
Different $P$ leads to different Higgs sector, corresponding to different particle spectrum and residual symmetry at low energy.
If there exists a special permutation $P_{1}$ such that $|z_{1}+z_{P_{1}(1)}| = |z_{2}+z_{P_{1}(2)}| = \dots = |z_{N}+z_{P_{1}(N)}|$, a sole Higgs singlet $S$ with charge $z_{S} = |z_{1} + z_{P_{1}(1)}|$ is sufficient to generate all masses.
To determine whether this happens to $\DF{z_{1}}, \dots, \DF{z_{N}}$, one needs to exhaust all possible permutation $P$.
If every $P$ generates Higgs sector containing more than one Higgs, introducing only one Higgs singlet would give rise massless state inevitably.

\bibliography{main}

\end{document}